\documentclass[11pt]{article}
\usepackage{graphicx}
\usepackage{amssymb}

\title{A realistic interpretation of quantum mechanics. Asymmetric random walks in a discrete spacetime.}
\author{Antonio Sciarretta}

\begin{document}
\maketitle
\section{Abstract}

In this paper, I propose a realistic interpretation (RI) of quantum mechanics, that is, an interpretation according to which a particle follows a definite path in spacetime. The path is not deterministic but it is rather a random walk. However, the probability of each step of the walk is found to depend from some average properties of the particle that can be interpreted as its propensity to have a certain macroscopic momentum and energy. The proposed interpretation requires spacetime to be discrete. Prediction of standard quantum mechanics coincide with predictions of large ensembles of particles in the RI.

\section{Introduction}

Despite its general acceptance as a tool to predict the outcome of experiments with particles and other microscopic objects, standard quantum mechanics (QM) is believed by many to be incomplete or, at least, not fully understood. In particular, the ``strange" or non-classical phenomena of QM, like self-interference and Born's rule, are described in terms of abstract mathematical objects. Although some proponents of the standard or Copenhagen intepretation of QM might have been ready to interpret the complex-valued wavefunction as a real object \cite{kiefer}, wavefunctions are essentially mathematical tools serving to calculate probabilities from their square moduli. Contrasting to real-valued mathematics and one-to-one mapping between real variables and observables of classical theories, the standard description is thus often considered as purely operational and thus not realistic. 

Consider the quantum phenomenon of particle self-interference, as illustrated by the double-slit experiment. In the standard picture, self-interference is due to the intrinsic superimposition of the two complex wavefunctions associated to the two slits. The wavefunction is believed to describe every single particle, so that a particle is said to pass simultaneously from both slits. That implies particle-wave duality, or other hardly imaginable concepts like diffusion in imaginary time \cite{nagasawa}. 

Proponents of alternative views have tried to re-conciliate QM with more realistic assumptions concerning particle behavior. A well-known example is constituted by hidden-variable theories like De Broglie's pilot wave theory or Bohmian mechanics \cite{bohm}, a more recent variant of which is the deterministic trajectory representation \cite{floyd}. These theories calculate actual particle trajectories \cite{philippidis} that accumulate or rarefact, leading to maxima and minima of fringes at the screen behind the slits. However, such an approach lacks a plausible mechanism for justifying non-localities, i.e., describing how the wave arises and does its guiding. 

The advocates of ensemble or statistical interpretation have recognized the standard description of QM as a description that does not apply to individual particles but rather to ensembles of similarly prepared particles \cite{ballentine, ballentine1}. However, ensemble interpretation does not provide any alternative way to explain the behavior of single particles, nor it fills the gap between the classical domain and the quantum domain \cite{neumaier}. 

To explain self-interference in a realistic manner several alternative approaches have been developed. A clearly non-exhaustive list includes theories involving the influence of the context \cite{khrennikov, khrennikov1}, nonstandard probability definitions \cite{youssef}, nonstandard types of particle trajectories \cite{qi}, quanta as real particles acting as sources of real waves \cite{mardari}, nonlocality and discrete spacetime \cite{santanna}. Despite these efforts, QM is still universally believed to lack a convincing realistic interpretation.

In this paper one possible realistic interpretation is presented. Although very simple in its mathematical development and formalism, the proposed approach seems to be able to predict self-interference and probability fields independently from Schr\"{o}dinger equation or wavefunctions. Unlike the standard picture, only real quantities (actually, integers) are employed. In particular, this approach assumes a discrete spacetime. Unlike ensemble interpretation, it describes single particle trajectories, however, as Markov chains, i.e., with an intrinsic randomness. Transition probabilities are simple functions of momentum propensity. The latter is randomly determined at the preparation phase. Consequently, probability distributions of similarly-prepared ensembles of particles are obtained in accord to standard QM predictions.

The paper is organized as follows. Section~1 presents the assumptions concerning spacetime, which naturally lead to a realistic interpretation of uncertainty principle. Section~2 provides a realistic interpretation of Schr\"{o}dinger equation in terms of Markov chain and transition probabilities. Section~3 presents the results for a self-interference case, with only one additional assumption concerning the probability density of momentum propensity as determined at the sources. Section~4 finally presents a particle-by-particle simulation of a double-source experiment, whose a posteriori distribution of arrival frequency practically tends to coincide with predictions of Schr\"{o}dinger equation.

\section{The lattice: uncertainty principle}

The RI assumes that the spacetime is inherently discrete. Limiting for simplicity the analysis to one dimension $x$, the RI assumes that only values $x=\xi X$, $\xi=0,1,2,\ldots$ and $t=\tau T$, $\tau=0,1,2,\ldots$ are meaningful. Noninteger values of space and time are simply impossible in this picture. The two fundamental quantities $X$ and $T$ are the size of the lattice that constitutes the space and the fundamental temporal resolution, respectively.

Under this assumption, a particle trajectory consists of a succession of points $\{\xi,\tau\}$ in the spacetime. Advance in time is unidirectional and unitary, that is, $\tau+1$ follows necessarily $\tau$. Advance in space is still unitary but bidirectional. If at a time $\tau$ a particle resides at the location $\xi$ of the spatial lattice, at time $\tau+1$ the particle can only reside at locations $\xi+1$, $\xi$, or $\xi-1$. In other words, the local velocity in lattice units $\beta=\Delta \xi/\Delta \tau$ can only take the values $+1$, $0$ or $-1$. 

One must distinguish between local velocity and average or macroscopic velocity of the particle. The meaning of the latter is intuitive. Its definition in lattice terms is
\begin{equation}
  v = \lim_{N\rightarrow \infty} \frac{1}{N} \sum_{\tau=1}^N \beta(\tau) \frac{X}{T} 
  \label{eqn:v}
\end{equation}

The maximum velocity that a particle can reach is the speed of light $c$. Light trajectory in the positive direction corresponds to $\beta(\tau)=1$, $\forall \tau$. Consequently to (\ref{eqn:v}), one constraint to the fundamental lattice quantities is necessarily
\begin{equation}
  \frac{X}{T} = c 
  \label{eqn:cc} 
\end{equation}

Another consequence of (\ref{eqn:v}) is that to determine the macroscopic velocity of a particle, an observer should wait in principle a time $N$ tending to infinity. Every observation lasting a finite amount $N$ of time steps will give an approximation of $v$. Consider, e.g., $N=1$. The observed velocity can be $+1$, $0$ or $-1$. Thus the uncertainty of the macroscopic velocity $v$ is 1 in absolute value. For $N=2$, the possible results for the observed velocity are $1$, $1/2$, $0$, $-1/2$, and $-1$. Thus the uncertainty of the macroscopic velocity is 1/2 in absolute value. Extending these considerations, the uncertainty of the macroscopic velocity after an observation lasting $N$ time steps is $1/N$ in lattice units.

Moreover, an observation lasting $N$ time steps necessarily implies a change in the position of the particle. The span of the particle during the observation ranges from $NX$ to $-NX$. Thus the uncertainty of the position of the particle at the end of the observation is obviously $2N$ in lattice units.

Using the two results above, and labelling $\Delta v(N)$ and $\Delta x(N)$ the uncertainties of velocity and position as a function of observation time $N$, the relationship
\begin{equation}
  \Delta v(N) \cdot \Delta x(N) = \frac{c}{N} \cdot 2NX = \frac{2X^2}{T}
  \label{eqn:heis}
\end{equation}
holds.

The latter equation resembles the Heisenberg uncertainty principle since it fixes an inverse proportionality between the uncertainty with which the velocity of a particle can be known and the uncertainty with which its position can be known. Multiplying by the particle mass $m$, and comparing (\ref{eqn:heis}) to Heisenberg uncertainty principle, one obtains that the two fundamental lattice quantities are related to the Planck constant,
\begin{equation}
  m \frac{X^2}{T} = \frac{\hbar}{2} = \frac{h}{4\pi}  
  \label{eqn:hbar}
\end{equation}
The term $4\pi$ (the solid angle of a sphere) is for three dimensional spaces. In our example case of a 1-d space, this term reduces to $\Omega_1=\frac{2\pi^{1/2}}{\Gamma(1/2)}$, where $\Gamma$ is the Gamma function, i.e., $\Omega_1=2$. Thus, combining (\ref{eqn:cc}) with the accordingly modified (\ref{eqn:hbar}), the values for the fundamental lattice quantities are obtained as 
\begin{equation}
  X=\frac{h}{2mc}
\end{equation}
and
\begin{equation}
  T = \frac{h}{2mc^2}
\end{equation}

These values clearly correspond to the Planck length and the Planck time, respectively. The role of mass is not completely clear at this point. Likely, general relativity might serve to integrate it into the picture.

\section{Propagation: equation of Schr\"{o}dinger}

Now I describe the propagation rules of a particle on the lattice in the RI. At each time $\tau$, the particle might jump to one of the nearest neighboring sites of the lattice, or stay at rest. The actual local trajectory is not deterministic, i.e., it is not a prescribed function of previous parts of trajectory. Rather, the local trajectory has the characteristics of a random walk. This point is very important and it implies that an intrinsic randomness affects the particle motion. However, in contrast to pure random walks, there is a different transition probability for each possible jump. Label the three probabilities $a$, $b$, and $c$, respectively. Of course, \begin{equation}
  a+b+c=1
  \label{eqn:abc}
\end{equation}
Moreover, the RI assumes that the average or macroscopic momentum, $p=v/c$ in lattice units (consider $m=1$ henceforth), is imprinted to the particle. This imprint is to be attributed to the preparation process. It is thus possible to talk about \emph{momentum propensity}. The quantity $p$ can be also reinterpreted as the probability of unitary motion in the positive direction, so that 
\begin{equation}
  a-c = p
  \label{eqn:p}
\end{equation}

Another average or macroscopic quantity is the energy of the particle, which in lattice units is written as 
\begin{equation}
  e = \lim_{N\rightarrow \infty} \frac{1}{N} \sum_{\tau=1}^N |\beta(\tau)| 
  \label{eqn:e}
\end{equation}
Also this energy can be reinterpreted as the propensity to unitary motion in either direction. Thus
\begin{equation}
  a+c = e
  \label{eqn:eps}
\end{equation}
Combining (\ref{eqn:abc})--(\ref{eqn:eps}), one obtains that
\begin{equation}
  a = \frac{e+p}{2}, \quad b = 1-e, \quad c = \frac{e-p}{2}
  \label{eqn:a}
\end{equation}

The energy $e$ must be a function of $p$. A well-known result of special relativity states that energy of a particle is the sum of the rest energy and the kinetic energy. In lattice units, that proposition can be written as
\begin{equation}
  e(p) = \frac{1+p^2}{2}
  \label{eqn:energy}
\end{equation}
Consequently, (\ref{eqn:a}) can be rewritten as
\begin{equation}
  a = \left(\frac{1+p}{2}\right)^2, \quad b = \frac{1-p^2}{2}, \quad c = \left(\frac{1-p}{2}\right)^2
  \label{eqn:a1}
\end{equation}

\begin{figure}[t!]
\centering
\includegraphics[width=.6\textwidth]{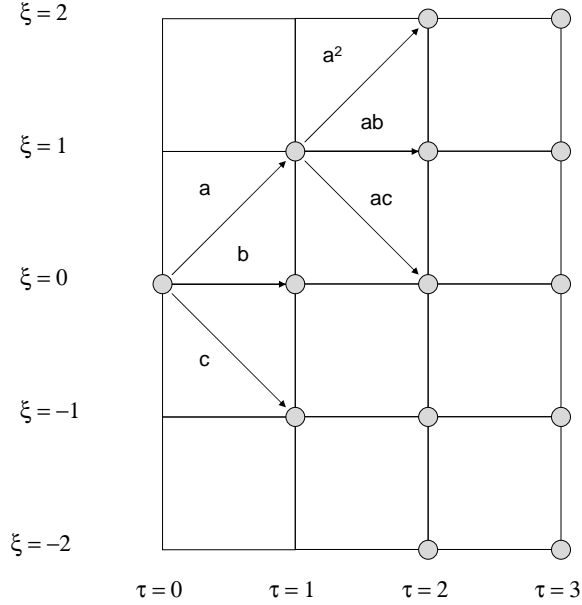}
\caption{Particle trajectories after emission from a single source. The graph illustrates the neighboring points on the lattice and the three elementary probabilities $a,b,c$.}
\label{fig:lattice}
\end{figure}

The equations above, in fixing the probability of each jump at each time step $\tau$, determines the trajectory of the particle as a random walk. For example, observe Fig.~\ref{fig:lattice}. A particle is emitted from a source located at the site $\xi=0$ of the lattice with an intrinsic value of $p$ and thus of $e$, determined by the preparation (I will return soon on this point). After one time step, $\tau=1$, the particle has a probability $a$ to be at the site $\xi=1$, a probability $b$ to be at the site $\xi=0$, a probability $c$ to be at the site $\xi=-1$. After two time steps, $\tau=2$, the probabilities for each site from $\xi=-2$ to $\xi=2$ are as follows: $\rho_p(2)=a^2$, $\rho_p(1)=2ab$, $\rho_p(0)=2ac+b^2$, $\rho_p(-1)=2bc$, $\rho_p(-2)=c^2$. Notice that, since the functions $a(p)$ and $c(p)$ are symmetric, the probability distribution is symmetric with respect to $\xi=0$. Notice also that $b^2=4ac$ and thus every probability $\rho_p$ is expressed as a unique combination of the three elementary probabilities, multiplied by a coefficient.

In general, the recursive expression for the probability of finding the particle at time $\tau$ at site $\xi$ is given by
\begin{equation}
  \rho_p(\xi,\tau) = a\rho_p(\xi-1,\tau-1)+b\rho_p(\xi,\tau-1)+c\rho_p(\xi+1,\tau-1)
  \label{eqn:process}
\end{equation}
Of course, the start of the recursion is $\rho_p(0,0)=1$ for the case of a single source. Deriving a closed formula for the probability $\rho_p(\xi,n)$ is tedious but straightforward at this point. The derivation conduces to binomials and binomial coefficients. Instead of attempting to present here the derivation, I will show numerical results in the following. However, a special case is easy to calculate in closed form. The probability that a particle is at the event horizon, i.e., $\rho_p(\tau,\tau)$, is easily calculated as $a^\tau$. Similarly, $\rho_p(-\tau,\tau)=c^\tau$.

Now, a delicate passage in the theory is introduced. The RI assumes that the momentum propensity $p$ is determined \emph{randomly} during the preparation at the source. The probability of releasing a particle with a momentum $p$ is uniform over the possible values of $p$. Since $p$ can vary between -1 and +1, its span is 2 and thus the probability density $f(p)=1/2$. 

If the source releases a large number of particles in succession, each one with a randomly determined value of $p$, the probability of finding a particle at the location ${\xi,\tau}$ is clearly given by
\begin{equation}
  \rho(\xi,\tau) = \int_{-1}^{1}f(p)\rho_p(\xi,\tau)dp = \frac{1}{2}\int_{-1}^{1}\rho_p(\xi,\tau)dp
  \label{eqn:intp}
\end{equation}
Let us calculate the integral of (\ref{eqn:intp}) for the special case mentioned above, that is,
\begin{equation}
  \rho(\tau,\tau) = \frac{1}{2}\int_{-1}^{1}a^\tau dp = \frac{1}{2}\int_{-1}^{1}\left(\frac{1+p}{2\tau}\right)^\tau dp
  \label{eqn:intpnn}
\end{equation}
Using some elementary mathematics,
\begin{equation}
  \rho(\tau,\tau) = \frac{1}{2^{2\tau+1}}\left[\frac{(1+p)^{2\tau+1}}{2\tau+1}\right]_{-1}^{1} = \frac{1}{2^{2\tau+1}}\frac{2^{2\tau+1}}{2\tau+1} = \frac{1}{2\tau+1}
  \label{eqn:intnn2}
\end{equation}
This result is not casual. It can be shown numerically that the same probability is valid for every other site location comprised between the two event horizons, i.e.,
\begin{equation}
  \rho(\xi,\tau) = \frac{1}{2\tau+1}, \forall \xi\in [-\tau,\tau]
  \label{eqn:rho}
\end{equation}
It is easily verified that 
\begin{equation}
  \sum_{\xi=-\tau}^{\xi=\tau} \rho(\xi,\tau) = 1
  \label{eqn:sumrho}
\end{equation}
 
Now, let us compare this result with the predictions of standard QM, i.e., the particular solution of the Schr\"{o}dinger equation. The wavefunction for a free particle is
\begin{equation}
  \Psi(x,t) = \frac{1}{\sqrt{2\pi}} \int e^{i(kx-\omega t)}\varphi(k,0)dk
  \label{eqn:psi}
\end{equation}
where the wavenumber $k$ is related to the momentum of the particle and
\begin{equation}
  \varphi(k,0) = \frac{1}{\sqrt{2\pi}} \int \Psi(x,0)e^{-ikx}dx
  \label{eqn:phi}
\end{equation}
For a single perfectly localized source at $x=0$, $\Psi(x,0)=\delta(x)$. Thus (\ref{eqn:phi}) reads
\begin{equation}
  \varphi(k,0) = \frac{1}{\sqrt{2\pi}}
  \label{eqn:phi0}
\end{equation}
and consequently (\ref{eqn:psi}) is rewritten as
\begin{equation}
  \Psi(x,t) = \frac{1}{\sqrt{2\pi}} \int \exp\left[i(kx-\frac{\hbar k^2}{2m} t)\right]dk
  \label{eqn:psi1}
\end{equation}
that, integrated, yields
\begin{equation}
  \Psi(x,t) = \frac{1}{\sqrt{2\pi}}\sqrt{\frac{m}{i\hbar t}}\exp\left[i\frac{mx^2}{2\hbar t}\right]
  \label{eqn:psi2}
\end{equation}
The probability density is easily calculated as
\begin{equation}
  |\Psi(x,t)|^2 = \frac{m}{2\pi\hbar t} = \frac{m}{ht}
  \label{eqn:f0}
\end{equation}
thus it is inversely proportional to time and it does not depend on $x$. Normalizing time to lattice units and using (\ref{eqn:hbar}) allows reducing (\ref{eqn:f0}) to
\begin{equation}
  \textrm{prob. density} = \frac{1}{2\tau}\frac{1}{X^2}
  \label{eqn:f1}
\end{equation}
The probability $P(x,t)=|\Psi(x,t)|^2 X^2$. The result compares with (\ref{eqn:intnn2}), with $2\tau$ replacing $2\tau+1$. The two functions of $\tau$ are very similar and, indeed, practically coincident for $\tau$ sufficiently large. That could be interpreted in the following way:
\begin{quote}
The square modulus of the wavefunction predicted by the Schr\"{o}dinger equation is an approximation of the probability in RI. This approximation is as better as the measurement is farther from the source. 
\end{quote}

Another interesting result arises from the process (\ref{eqn:process}). Let us calculate the \emph{action} of the particle, i.e., the energy accumulated by the particle during its walk, $\sigma(\xi,\tau)$. Let us introduce also the variable $\Sigma(\xi,\tau)=\sigma(\xi,\tau)/\rho(\xi,\tau)$. The random process for $\sigma$ is given by
\begin{equation}
  \sigma(\xi,\tau) = a(\Sigma(\xi-1,\tau-1)+1)+b\Sigma(\xi,\tau-1)+c(\Sigma(\xi+1,\tau-1)+1)
  \label{eqn:sigma}
\end{equation}
with the initial condition $\Sigma(0,0)=0$. 

For example, one easily obtains $\sigma(1,1)=a$ and thus $\Sigma(1,1)=1$. The action of every particle reaching the point $\{1,1\}$ is obviously 1. Generalizing this result, clearly $\sigma(\tau,\tau)=\tau a^2$ and thus $\Sigma(\tau,\tau)=\tau$. For a point like $\{2,0\}$ the prediction is less trivial. The possible values of action can be 2, if the particle follows a back-and-forth path, or 0, if it stays at rest for two time steps. Using (\ref{eqn:sigma}), one obtains for this case $\sigma(2,0)=4ac$ and thus $\Sigma(2,0)=2/3$, which is a weighted mean between the two possible values of action. 

The results easily calculated for other points are listed in Table~\ref{tab:1}. Observing the trend of $\Sigma$ for a given $\tau$, one discovers a clear quadratic dependency on $\xi$. Indeed, the function $\Sigma(\xi,\tau)$ is calculated as
\begin{equation}
  \Sigma(\xi,\tau) = \frac{\xi^2+\tau^2-\tau}{2\tau-1} = \Sigma_0(\tau)+\frac{\xi^2}{2\tau-1}
  \label{eqn:s}
\end{equation}
where $\Sigma_0$ is the action at $\xi=0$. Equation (\ref{eqn:s}) can be easily verified by inspection of a few points, provided that $\{\xi,\tau\}>0$. 

Now, compare this result with the phase of the wavefunction (\ref{eqn:psi2}). The latter, usually interpreted as the action of the particle is
\begin{equation}
  S(x,t)=\frac{mx^2}{2\hbar t}
\end{equation}
which, in lattice units, is
\begin{equation}
  S(x,t)=2\pi\frac{\xi^2}{2\tau}
\end{equation}
The latter equation corresponds to the second terms in the right-hand side of (\ref{eqn:s}), that is, $\Sigma-\Sigma_0$, multiplied by $2\pi$ to obtain a phase angle. The correspondence is almost perfect, except for the term $2\tau-1$ that in the RI replaces the term $2\tau$ predicted by standard QM. For large values of $\tau$, however, the two results are practically coincident. In other terms: 
\begin{quote}
The phase predicted by the complex Schr\"{o}dinger equation is an approximation of the action in RI. This approximation is as better as the measurement is farther from the source. 
\end{quote}

\begin{table}[t!]
\caption{Calculated values of $\Sigma(\xi,\tau)$ for some small values of $\xi$ and $\tau$.}
\begin{center}
\includegraphics[width=.6\textwidth]{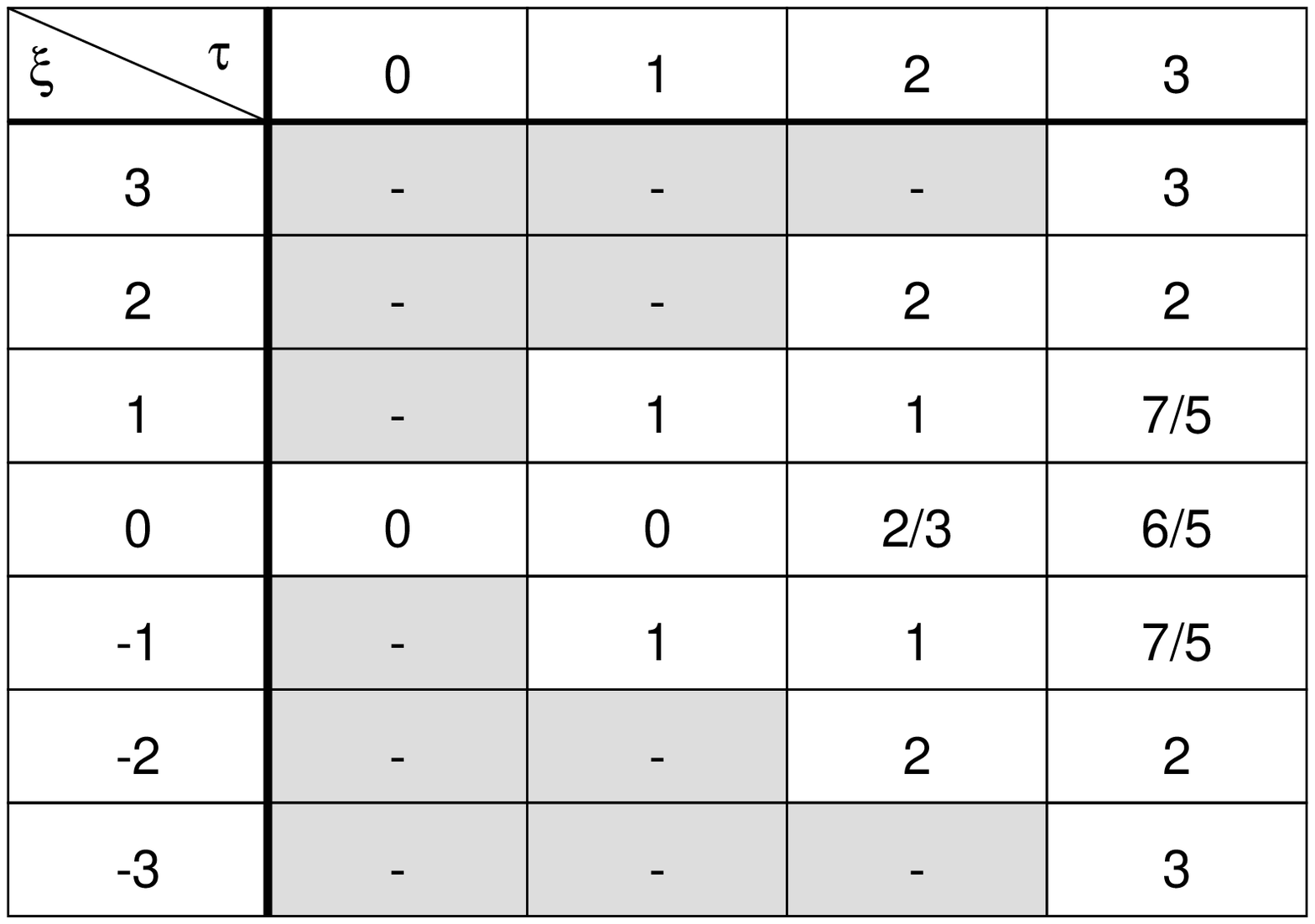}
\end{center}
\label{tab:1}
\end{table}

In addition to probability of arrival and action, the source determines the whole state of the particle, that is, other observables. In order to calculate a few of them, I find useful to introduce global probability fluxes, which are the probability of each possible path (``history", in standard QM language) summed over all possible $p$'s. The flux $J^{\{+,0,-\}}(\xi,\tau)$ describes the probability of jumping to site $\xi$ from site $\xi-1$, $\xi$, and $\xi+1$, respectively, at time $\tau$. 

In terms of global probability fluxes, the probability and the action at a certain site are also calculated as
\begin{equation}
  \rho(\xi,\tau) = J^+(\xi,\tau)+J^0(\xi,\tau)+J^-(\xi,\tau)  \label{eqn:rhon}
\end{equation}
\begin{equation}
\begin{array}{rcl}
  \Sigma(\xi,\tau) & = & \frac{1}{\rho(\xi,\tau)}\left\{J^+(\xi,\tau)[\Sigma(\xi-1,\tau-1)+1]+ \right. \\
& & \left. +J^0(\xi,\tau)[\Sigma(\xi,\tau-1)]+J^-(\xi,\tau)[\Sigma(\xi+1,\tau-1)+1]\right\}
\end{array}
  \label{eqn:action}
\end{equation}
Other observables of the random walk that can be calculated with this approach are, for instance, the average momentum (the actual observable momentum, not the momentum propensity that is a property of the particle) at a certain location of the lattice, which is given by 
\begin{equation}
  \bar{p}(\xi,\tau) = \frac{J^+(\xi,\tau)-J^-(\xi,\tau)} {J^+(\xi,\tau)+J^0(\xi,\tau)+J^-(\xi,\tau)}
  \label{eqn:pbar0}
\end{equation}
and is calculated as
\begin{equation}
  \bar{p} = \frac{\xi}{\tau}
  \label{eqn:pbar}
\end{equation}
Similarly, the average energy at a certain location is given by
\begin{equation}
  \bar{e}(\xi,\tau) = \frac{J^+(\xi,\tau)+J^-(\xi,\tau)} {J^+(\xi,\tau)+J^0(\xi,\tau)+J^-(\xi,\tau)}
\end{equation}
and it is calculated as 
\begin{equation}
  \bar{e} = \frac{\xi^2}{\tau(2\tau-1)}
  \label{eqn:ebar}
\end{equation}
Both the latter equation have a counterpart in the variables predicted by the Schr\"{o}dinger equation, in particular in the Bohm interpretation. The local average momentum corresponds to the momentum of Bohm hidden variable, 
\begin{equation}
  \nabla S = \frac{mx}{\hbar t}
  \label{eqn:vflux}
\end{equation}
so that the probability flux is $j=\frac{\rho}{m} \nabla S$. The local average energy corresponds to the energy of Bohm hidden variable, that is, time derivative of the action
\begin{equation}
  -\frac{\partial S}{\partial t} = \frac{mx^2}{2\hbar t^2} = \frac{1}{2m}(\nabla S)^2
  \label{eqn:st}
\end{equation}
To see the correspondence it is sufficient to write (\ref{eqn:vflux})--(\ref{eqn:st}) in lattice units or, alternatively, to put $m=1$, $\hbar=1$ in both of them. As for the other observables analyzed above, the correspondence is better as $\tau$ increases.

\section{Multiple sources: interference}

\begin{figure}[t!]
  \centering
  \includegraphics[width=.6\textwidth]{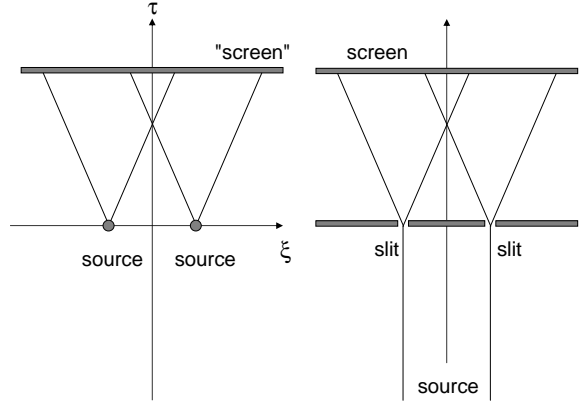}
  \caption{Pictorial equivalence between a double-slit experiment in two dimensions and a double-source test case in one dimension.}
  \label{fig:slits}
\end{figure}

After having reproduced the predictions of the Schr\"{o}dinger equation for a free particle, I proceed now to a second puzzling aspect of QM: particle self-interference. Double-slit experiment usually serves to visualize this phenomenon. However, the core of  self-interference is isolated and better illustrated by a double-source preparation. Instead of having a single source, a two-slit barrier, and a screen behind the barrier, I represent the same process with two independent and mutually alternative sources of particles, separated by a certain distance $2dX$, and a screen. The sources are equivalent to very narrow, i.e., punctiform slits. Being in a one-dimensional space, the location of the ``screen" is clearly fictitious. Pictorially, the geometry of the system can be still imagined in two dimensions. One dimension is $\xi$, along which the particle move with a momentum propensity $p$. The second dimension is perpendicular to $\xi$ and is traversed by the particle with momentum propensity 1 (certainty of advancing in the positive direction). The ``screen" is thus located at a distance $\tau$ from the sources. Figure~\ref{fig:slits} illustrates this equivalence.

The solution of the Schr\"{o}dinger equation for this case is based on the linear superimposition of the two waveforms relative to the two sources, 
\begin{equation}
  \Psi(x,t) = \frac{\Psi_1(x,t)+\Psi_2(x,t)}{\sqrt{2}}
\end{equation}
where both $\Psi_1(x,t)$ and $\Psi_2(x,t)$ are obtained from (\ref{eqn:psi2}) by replacing  the term $x^2$ (that was valid for a source at $x=0$) with a term $(x-dX)^2$ and $(x+dX)^2$, respectively. Thus
\begin{equation}
  \Psi(x,t) = \frac{1}{2\sqrt{\pi}}\sqrt{\frac{m}{i\hbar t}}\left\{\exp\left[i\frac{m(x-dX)^2}{2\hbar t}\right]+\exp\left[i\frac{m(x+dX)^2}{2\hbar t}\right]\right\}
  \label{eqn:2psi}
\end{equation}
The probability density is given by
\begin{equation}
  |\Psi(x,t)|^2 = \frac{2m}{h t}\cos^2\left(\frac{S_1-S_2}{2}\right)
\end{equation}
where $S_1$ and $S_2$ are the two independent action values, that is, the phases of the two exponentials in (\ref{eqn:2psi}). Finally, the probability in lattice units is
\begin{equation}
  P(\xi,\tau) = \frac{1}{\tau}\cos^2\left(\pi\frac{\left(\xi-d\right)^2-\left(\xi+d\right)^2}{2\tau}\right)
  \label{eqn:p2slit}
\end{equation}
and thus an interference term arises due to the presence of two possible sources. The interference is related to the phase difference between the two waveforms.

The representation of the same process in the RI, i.e., in terms of discrete spacetime and random walk, would give just the superimposition of two probability densities of the type (\ref{eqn:rho}), if the same initial conditions are taken as in the single-source case. No interference term would arise in this case.

The key factor to represent self-interference resides in the choice of the correct probability density of the momentum propensity $p$. For a two source process, $f(p)\neq 1/2$. Some values of momentum are more probably than others. This fact may seem strange, but actually it is already contained in the standard picture of QM \cite{marcella}. To verify it, it is sufficient to apply (\ref{eqn:phi}) with $\Psi(x,0)=(\delta(x-dX)+\delta(x+dX))/\sqrt{2}$. The result is
\begin{equation}
  \varphi(k,0) = \frac{1}{\sqrt{2\pi}}\left(e^{-ikd}+e^{ikd}\right)
\end{equation}
from which the probability function is
\begin{equation}
  f(k) = \frac{1+\cos(2dk)}{2\pi}
  \label{eqn:fk}
\end{equation}
Recalling that $k=p/\hbar$, (\ref{eqn:fk}) is transformed in lattice units as
\begin{equation}
  f(p) = \frac{1+\cos(4\pi d p)}{2}
  \label{eqn:fp}
\end{equation}
which is a function oscillating between 1 and 0 with a mean value of 1/2.

Again, the fact that some momenta are more probable than others should not surprise. It is exactly what happens when the two sources are approximated by a barrier with two slits at a sufficiently large distance from a single source. The two slits filter the momenta spectrum, e.g., favoring the values corresponding to the directions of the axes that rely the source to the slits. The same behavior is apparently contained in the assumption (\ref{eqn:fp}).

Now, simply introducing (\ref{eqn:fp}) into (\ref{eqn:intp}), while still using the same process (\ref{eqn:process}) leading to $\rho_p$, describes the insurgence of self-interference. This is shown numerically in the next section. 

\section{Realistic simulation of particle trajectories}

In the last two sections, the predictions of the RI were shown in closed form, using mathematical equations in terms of \emph{a priori} probabilities and probability fluxes. The probability of a number of observable were calculated and found to be in accord to the predictions of standard quantum mechanics. Now, I will present numerical simulations of the random walk of \emph{single} particles and I will calculate the \emph{a posteriori} probabilities as frequencies over a large number of emissions. Thus, this section is aimed at reproducing numerically a true experiment. 

Table~\ref{tab:pseudo} shows the pseudocode used for such simulations. The two for-cycles are for the successively released $N_P$ particles, and for time up to $N_T$, which corresponds to the distance between the sources and the screen in the fictitious dimension perpendicular to $\xi$ (see discussion above). Each particle experiences the choice of three randomly-selected values: (i) the momentum propensity according to $f(p)$, (ii) the source from which it is emitted, and (iii) at each time step, its local velocity according to $p$. The final code line represents the counting of the particle that arrive at a certain location at time $N_T$. From this number of arrivals, an \emph{a posteriori} frequency $\nu(\xi)$ is calculated as the ratio to the total number of particles emitted.

\begin{table}
\caption{Pseudocode used for the simulations of Fig.~\ref{fig:res1}.} \vspace{.4cm}
\centering
\begin{tabular}{l} \hline
$i = 0$ \\
\textbf{for} particle 1 to $N_P$ \\
\hspace{1cm} $i = i+1$ \\
\hspace{1cm} $p(i)$ = random value beween -1 and +1 with prob. den. given by (\ref{eqn:fp}) \\
\hspace{1cm} $\xi(0)$ = random value +1 or -1 with probability 1/2 \\
\hspace{1cm} $s(0)$ = 0 \\
\hspace{1cm} $n$ = 0 \\
\hspace{1cm} \textbf{for} time 1 to $N_T$ \\
\hspace{1cm} \hspace{1cm} $n = n+1$ \\
\hspace{1cm} \hspace{1cm} $v$ = random value +1, 0 or -1 with prob. given by (\ref{eqn:a}) \\
\hspace{1cm} \hspace{1cm} $\xi(n)$ = $\xi(n-1)+v$ \\
\hspace{1cm} \hspace{1cm} $s(n) = s(n-1)+|v|$ \\
\hspace{1cm} \textbf{end for} \\
\hspace{1cm} $\nu(\xi(N_T)) = \nu(\xi(N_T))+1$ \\
\textbf{end for} \\ 
$\nu(\xi(N_T)) = \nu(\xi(N_T))/N_P$ \\ \hline
\end{tabular}
\label{tab:pseudo}
\end{table}

\begin{figure}[t!]
  \centering
  \includegraphics[width=\textwidth]{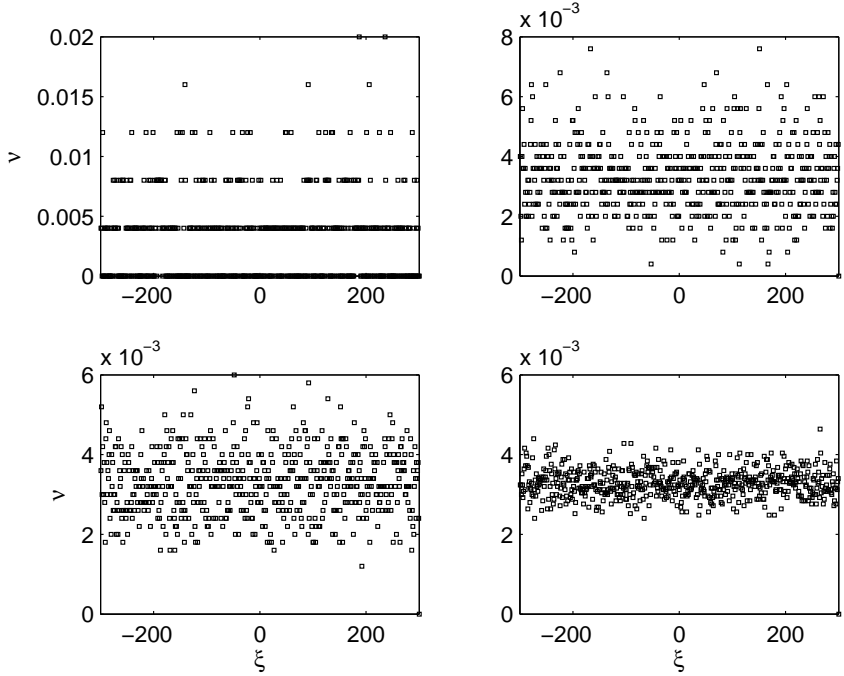}
  \caption{Probability of arrival of a particle emitted at $\xi=0$ as a function of $\xi$ after $N_T=300$. From top-left to bottom-right, $N_P=$ 500, 5000, 10000, and 50000, respectively.}
  \label{fig:buildprob}
\end{figure}

First, I present the case with only one source at $\xi=0$. Figure~\ref{fig:buildprob} shows the frequency $\nu(\xi)$ after a time $N_T=300$ for different values of $N_P$. As the the number of particles emitted in the ensemble increases, a frequency distribution builds up. For large $N_P$, the frequency clearly tends to the a priori probability $\rho(\xi,\tau=N_T)$, that is, a constant value given by (\ref{eqn:rho}).

\begin{figure}[t!]
  \centering
  \includegraphics[width=\textwidth]{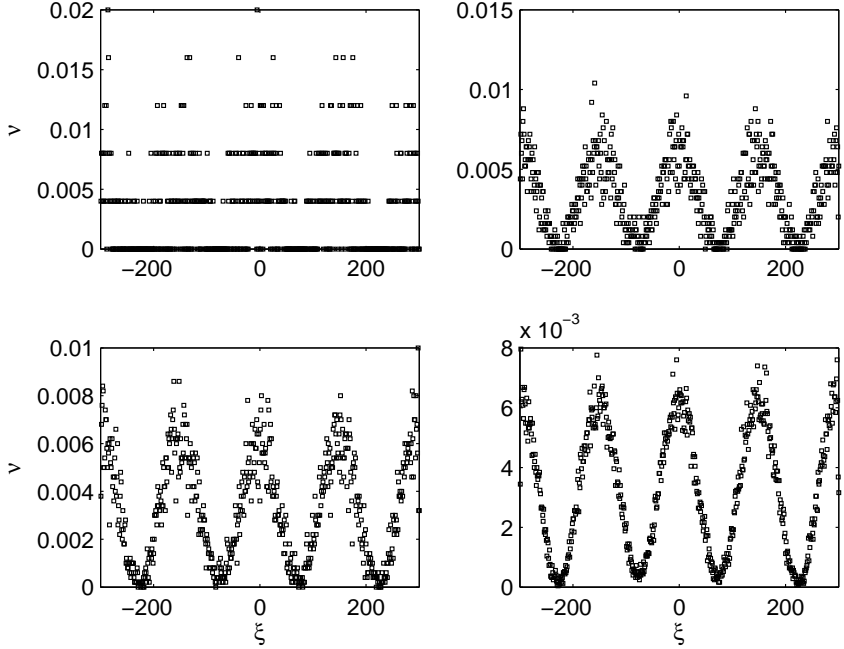}
  \caption{Probability of arrival of a particle emitted at $\xi=\pm 1$ as a function of $\xi$ after $N_T=300$. From top-left to bottom-right, $N_P=$ 500, 5000, 10000, and 50000, respectively.}
  \label{fig:buildinterf}
\end{figure}

The case with two sources at $\xi=\pm 1$ exhibit a completely different behavior, illustrated by Fig.~\ref{fig:buildinterf}. As the the number of particles emitted in the ensemble increases, an interference pattern clearly builds up. The simulations thus reproduce the outcome of famous experiments like those of Merli \emph{et al.} \cite{merli} and Tonomura \emph{et al.} \cite{tonomura}. Notice that, despite the fact that none of the probability fluxes of the RI can be negative or zero, nevertheless there are valleys in the function $\nu(\xi)$, that is, points where the particles seldom arrive.

\begin{figure}[t!]
  \centering
  \includegraphics[width=\textwidth]{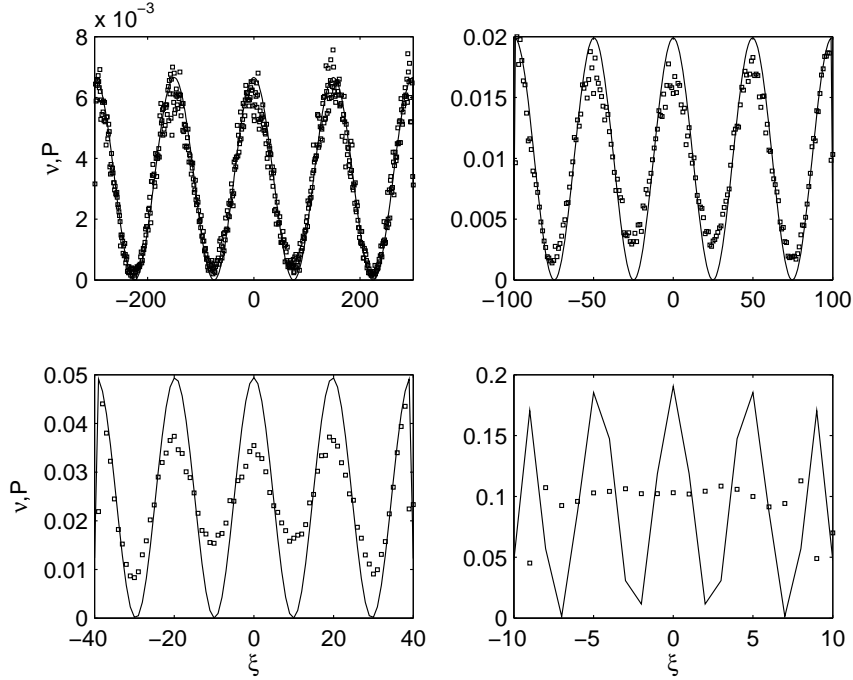}
  \caption{Probability of arrival of a particle emitted at $\xi=0$ as a function of $\xi$ after four values of $N_T$. From top-left to bottom-right, $N_T=$ 300, 100, 40, and 10, respectively. Solid lines: predictions with standard QM, $P(\xi,\tau=N_T)$. Squares: a posteriori probability $\nu(\xi)$ with the RI.}
  \label{fig:res1}
\end{figure}

Figure~\ref{fig:res1} shows how the frequency $\nu(\xi)$ varies with the observation time $N_T$. The number of particles is $N_P=50000$ in these simulations. The curves used as a basis of comparison are the predictions of standard QM, namely, the probability function $P(\xi,\tau=N_T)$ given by (\ref{eqn:p2slit}). 
Clearly, the latter approximate $\nu(\xi)$ as better as the observation time $N_T$ is longer. In the case $N_T=300$ the function $\nu(\xi)$ is very well smoothed by the analytical function $P(\xi,\tau=N_T)$. 

\section{Conclusions}

In contrast to the standard picture of QM, the RI does not only predict the probability of an event over an ensemble of similiarly prepared particles. Instead, it also represents a possible behavior of every single particle in the ensemble. Probability distribution of the various observables in the state are thus obtained \emph{a posteriori} by evaluating the relative frequency of occurrence of a certain event. Probability distributions can be also determined \emph{a priori}, by applying simple probability conservation equations. Moreover, while standard QM calculates the a priori probabilities as squared moduli of complex-valued wavefunctions, the RI directly yields the real-valued (actually, rational-valued) probabilities, without appealing to mathematical abstractions like complex numbers. 

Since no other proposals to describe in a realistic way the beavior of single particles exist so far, at least to the knowledge of the author, we might be tempted to believe that particles \emph{really} behave like described in the RI.

The results of this paper have concerned free particles. However, the RI seems naturally capable to integrate also external forces into the picture. Each interaction of the particle with its sourrounding is indeed expected to modify its intrinsic properties, namely, its momentum propensity.

\end{document}